\documentclass[fleqn,usenatbib]{mnras}

\usepackage{newtxtext,newtxmath}

\usepackage[T1]{fontenc}

\DeclareRobustCommand{\VAN}[3]{#2}
\let\VANthebibliography\thebibliography
\def\thebibliography{\DeclareRobustCommand{\VAN}[3]{##3}\VANthebibliography}

\usepackage{graphicx}	
\usepackage{amsmath}	
\usepackage{subcaption}
\usepackage{physics}

\newcommand{\D}{\text{d}}

\title[Charged Particle Spectra for Kilonova]{Spectra and Ionization Efficiencies of Charged Decay Particles in Kilonova Ejecta}

\author[van den Berg et al.]{
Levi N. van den Berg,$^{1,2,3}$\thanks{E-mail: l.n.vandenberg@kusastro.kyoto-u.ac.jp}
Kenta Hotokezaka,$^{2}$\thanks{E-mail: kentah@resceu.s.u-tokyo.ac.jp}
\\
$^{1}$Institute for Theoretical Physics, Utrecht University, Princetonplein 5, 3584 CC Utrecht, The Netherlands\\
$^{2}$Research Center for the Early Universe (RESCEU), Graduate School of Science,
The University of Tokyo, 7-3-1 Hongo, Bunkyo, Tokyo 113-0033, Japan\\
$^{3}$Department of Astronomy, Kyoto University, Kitashirakawa-Oiwake-cho, Sakyo-ku, Kyoto 606-8502, Japan
}

\date{Accepted XXX. Received YYY; in original form ZZZ}

\pubyear{\the\year{}}

\begin{document}
\label{firstpage}
\pagerange{\pageref{firstpage}--\pageref{lastpage}}
\maketitle

\begin{abstract}
Ionization by radioactive decay products including $\alpha$-particles, $\beta$-decay electrons, and fission fragments  plays a central role in determining the nebular-phase ionization state and spectra of kilonovae. 
In this work, ionization cross sections, stopping powers, thermalization histories, and particle degradation spectra are calculated self-consistently for $\alpha$-particles and fission fragments propagating through expanding kilonova ejecta. The treatment includes interactions with bound and free electrons, charge evolution of fission fragments, adiabatic energy losses, and particle spectra obtained from the continuous slowing-down approximation and Spencer--Fano formalism. 
The work per ion pair is evaluated for a range of ejecta compositions and ionization states. Despite the large differences in mass, charge, and injection energy between $\beta$-electrons, $\alpha$-particles, and fission fragments, the resulting work per ion pair is found to be remarkably similar across all decay channels and target species considered. In particular, heavy-element ions  exhibit nearly identical normalized ionization efficiencies for all decay products. This robustness arises because the ionization cross sections and stopping powers are governed by the same underlying collision physics, causing the ratio between ionization and energy loss to remain approximately constant over the relevant energy range. The results imply that the ionization state of late-time kilonova ejecta depends only weakly on the dominant radioactive decay channel, even in ejecta where $\alpha$-decay and fission dominate the heating budget.
\end{abstract}




\section{Introduction}
Understanding the cosmic origin of the chemical elements requires studying astrophysical environments capable of producing nuclei heavier than lithium, which primordial nucleosynthesis cannot form \cite{Peebles1966, Wagoner1967}. Neutron‑star mergers, observed as kilonovae, are leading candidates for sites of rapid neutron capture (the r‑process, \citealt{Burbidge1957,Cameron1957,Lattimer1974ApJ} already hinted at by \citealt{Alpher1948}). Ejecta from these mergers could consist of components spanning a wide range of electron fractions ($Y_e$) and velocities \citep[e.g.,][]{Radice2018ApJ,Fujibayashi2023ApJ} 
that should give rise to different observational signatures \citep[e.g.,][]{Kasen2017,Tanaka2018ApJ,kawaguchi2024}.

Spectroscopic study of the only gravitational‑wave identified kilonova to date (AT2017gfo/GW170817) has revealed production of elements up to the light lanthanides and several trans‑iron species (He\,I, Sr\,II, Y\,II, Te\,III, IV, La\,III, Ce\,III, in \citealt{Watson2019Natur.574..497W,Domoto2022,Hotokezaka2023Te, Sneppen2023, Sneppen2024A&AHe,Sneppen2024A&A,Tanaka2023,Tarumi2023,Mulholland2026MNRAS,chiba2026arXiv260405703C,Arya2026arXiv260405812A}), yet no unambiguous detection of actinides or other very heavy r‑process products has been reported. Predictions for abundances beyond atomic numbers $Z \gtrsim 98$ remain highly uncertain because of nuclear physics and mass‑model differences that can change yields by orders of magnitude (for an overview, see \cite{Holmbeck2023}), while the lower‑mass r‑process pattern is constrained by observations of metal‑poor stars \citep{Burris2000}. Signatures from long‑lived heavy isotopes, for example spontaneous fission of $^{254}$Cf, can substantially modify kilonova light curves at tens to hundreds of days; detecting such signatures would provide direct evidence for synthesis of superheavy nuclei in merger ejecta \citep{Wu2019PhRvL,Zhu2018}. Measurements of radioactive isotopes in meteoritic and geological samples, such as Pu-244 and Cm‑247, \citep{Wallner2015NatCo,Wallner2021Sci,Tissot2016} further indicate r‑process activity before solar‑system formation and impose constraints on candidate astrophysical sites  and timescales \citep{Hotokezaka2015NatPh,Bartos2019Natur.569...85B,Cote2019ApJ...878..156C,Wehmeyer2023ApJ,Chiesa2024ApJ}.

Previous studies of kilonova heating and late‑time light curves have examined the energy deposition and thermalization of beta decay, alpha decay and fission, demonstrating that alpha decay and fission can dominate heating on $\sim$10–100 day timescales \citep{Barnes2016,Hotokezaka2016MNRAS,Zhu2018,Wanajo2018ApJ,Waxman2019ApJ,Wu2019PhRvL,Hotokezaka2020}. Radioactive heating affects not only the kilonova luminosity but also the ionization state of the ejecta, which in turns determines the colour evolution and strengths of absorption and emission lines \citep[e.g.,][]{Hotokezaka2021,Pognan2023MNRAS,Brethauer2026ApJ,Jerkstrand2026MNRAS}.

However, treatments of ionization and spectral effects have typically adopted ionization rates calibrated to beta decay products rather than to the heavier, more energetic particles produced by alpha decay and fission. \textcolor{black}{The focus on beta decay is motivated by previous studies suggesting that beta-decay products dominate the heating of kilonovae at late times (e.g. \cite{Waxman2019ApJ}).} Beta electrons carry decay energies that scale with parent lifetimes, while $\alpha$-particles and fission fragments are born with kinetic energies determined by quantum tunnelling and fragment dynamics. The energy scales relevant are a few MeV per nucleus for $\alpha$-particles and tens to hundreds of MeV for fission fragments \citep{Gamow1928}. These higher energies, distinct stopping behaviours and different secondary cascade processes may change how decay energy thermalizes and how ionization proceeds within the ejecta.

This work derives ionization efficiencies, stopping powers and particle spectra tailored specifically to $\alpha$-particles and fission fragments propagating in kilonova ejecta, accounting for their characteristic energies and interaction channels, including ionization of bound electrons, nuclear scattering and collective interactions with free electrons\footnote{This all happens under the assumption that both fission and $\alpha$-particles are magnetically trapped, which seems to be the case as their respective Larmor radii are approximately $10^{13}$cm and $10^{12}$cm, for a relatively weak magnetic field of $0.1\mu$B at 10 days, where the ejecta radius is $10^{16}$cm \citep{Waxman2019ApJ}.}.
We couple these particle‑specific thermalization rates to the late‑time \textcolor{black}{heating rate, which we define as the energy deposited by the decay particles into the ejecta; in other words, the fraction of the decay energy that is not lost through the expansion of the medium. Quantifying how the partitioning of ionization changes when alpha particles and fission fragments are included in a self-consistent treatment of thermalization and ionization, compared with models that consider only beta-decay products.} 
Using the degradation spectra for a given stopping medium, we compute the energy required to create an electron-ion pair referred to as the work per ion pair.  We find that, for a plasma composed of r-process elements, the work per ion pair  for $\beta$ particles, $\alpha$ particles, and fission fragments agrees to within $<50\%$, demonstrating the robustness of the work per ion pair in the kilonova context. Furthermore, we provide an approximate formula to compute the work per ion for different ions.
In Section \ref{sec:therm} we develop the self-consistent framework needed to compute stopping powers and ionization cross sections for $\alpha$-particles and fission fragments, and use it to evaluate their thermalization. Section \ref{sec:ionization} introduces a prescription for their ionization efficiency and presents an analytic form for their instantaneous spectra. Section \ref{BetaIonSection} extends this treatment to $\beta$-electrons by providing their spectra and ionization cross sections, enabling a direct and comprehensive comparison of ionization efficiencies across all charged decay products in neutron-star merger ejecta in section \ref{sec:resultsanddiscussion}.

\section{Thermalization}\label{sec:therm}
\begin{figure*}
    \centering
    \begin{subfigure}[t]{0.49\textwidth}
        \centering
        \includegraphics[width=\textwidth]{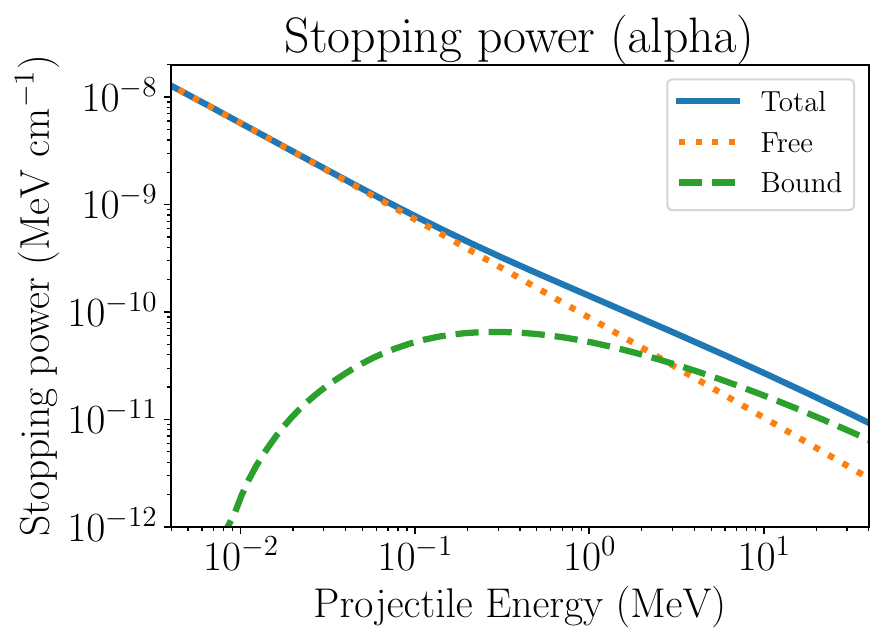}
        \label{fig:AlphaStopping}
    \end{subfigure}
    \hfill
    \begin{subfigure}[t]{0.49\textwidth}
        \centering
        \includegraphics[width=\textwidth]{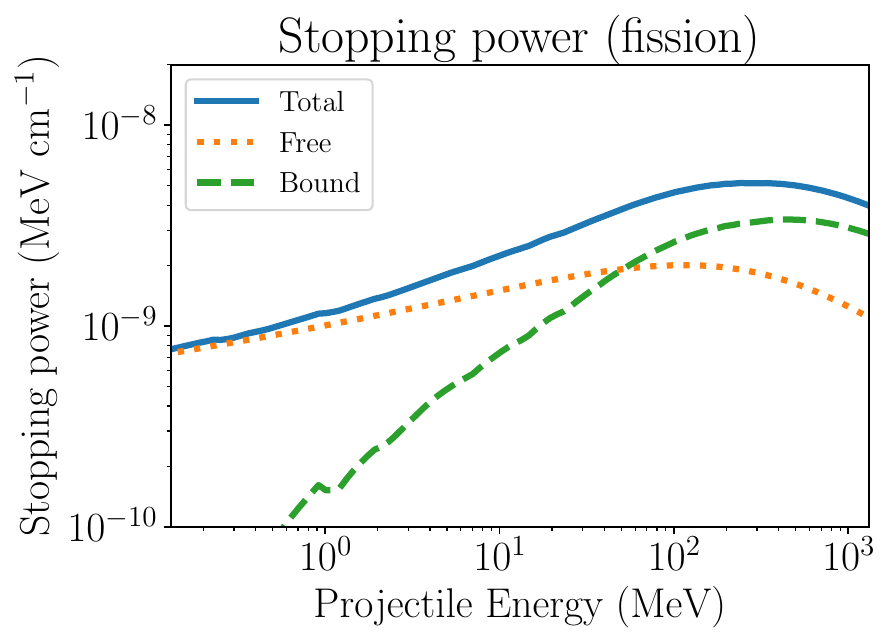}
        \label{fig:FissionStopping}
    \end{subfigure}
    \caption{Stopping power contributions from free and bound electrons for xenon fission fragments and $\alpha$-particles propagating through a Xe\,II plasma with $n_e = n_i = 1.19\times10^{9}\,\mathrm{cm^{-3}}$. For the fission fragment, the effective charge $Z_{\mathrm{eff}}$ is taken from the equilibrium values given by Eq.~\ref{SchiwietzFitFormula}, while $Z_{\mathrm{eff}} = 2$ is adopted for the $\alpha$-particles.}
    \label{fig:combined}
\end{figure*}

\begin{figure*}
	\centering
	\begin{minipage}[b]{0.49\textwidth}
		\centering
		\includegraphics[width=\linewidth]{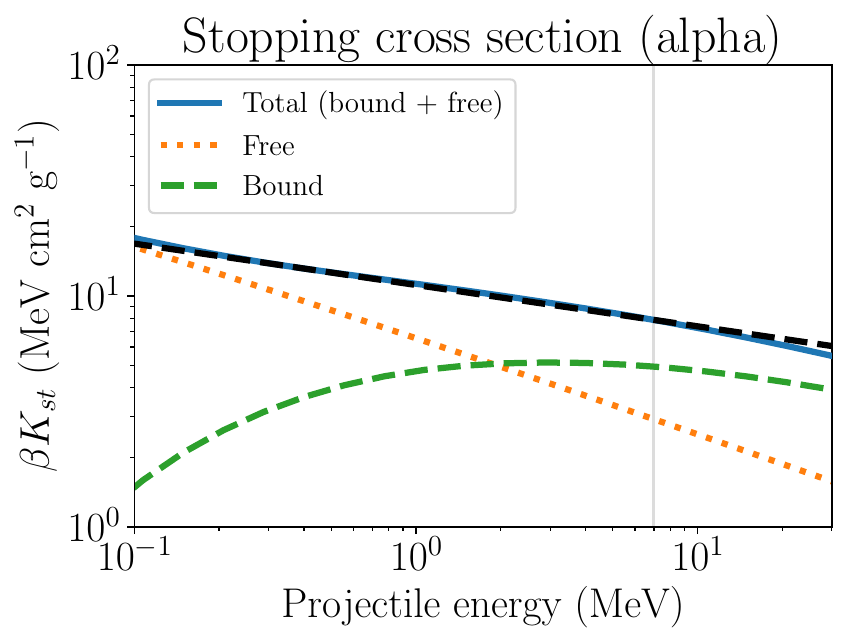}
	\end{minipage}
    \hfil %
	\begin{minipage}[b]{0.49\textwidth}
		\centering
		\includegraphics[width=\linewidth]{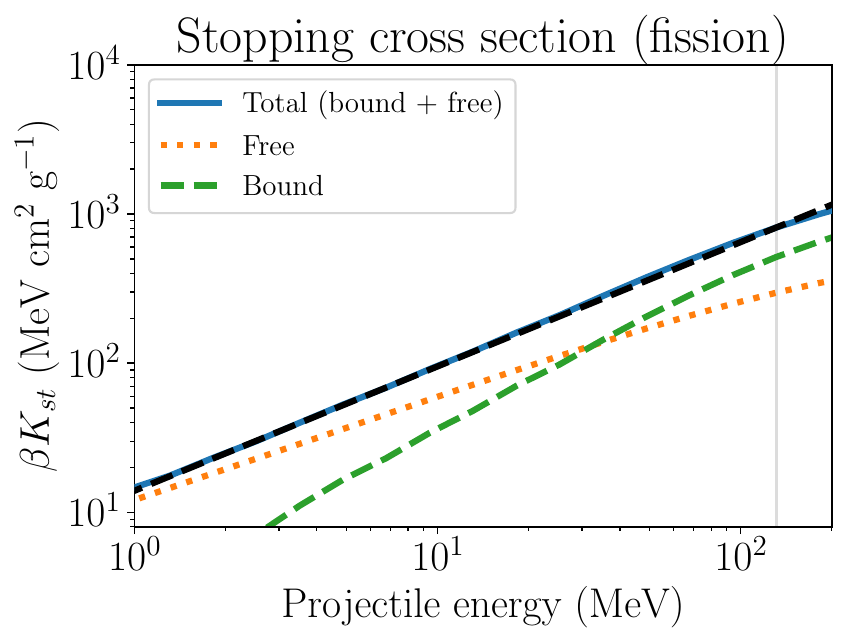}
	\end{minipage}
	\caption{Same configuration as in Fig. \ref{fig:combined}. The stopping cross sections $\beta K_{st}$, with $\beta = v/c$. An additional dashed black line has been drawn to illustrate the power law behaviour $\propto E^{\alpha}$, with $\alpha\approx 1.0$ for fission, and $\alpha\approx -0.2$ for $\alpha$-particles. Typical injection energies for fission and $\alpha$-particles have been indicated with a gray line at $E_0 \approx 130$ and $E_0 \approx 7$ MeV for fission and alpha respectively.}
	\label{fig:StoppingCrossSectionFissionAlpha}
\end{figure*}
Thermalization of decay products plays a critical role in determining the energy budget and ionization state of the kilonova ejecta. While the heating rate quantifies the initial energy deposition from fission and alpha decay fragments, not all of this energy contributes directly to ionization. The alpha and fission decay products lose their initial kinetic energy in  four channels: adiabatic expansion of the medium, collisions with free plasma electrons, collisions with electrons bound to plasma nuclei, and collisions with plasma nuclei themselves. Only the latter three are part of the thermalization. In addition, \textcolor{black}{energy losses through collisions with plasma nuclei are negligible because Coulomb repulsion between the positively charged decay products and target nuclei suppresses close encounters, making electronic stopping the dominant energy-loss mechanism (e.g. Chapter 2 in \cite{Sigmund2006}).} The relevant two channels are discussed below.

\subsection{Stopping Power} \label{subsec:stoppingpower}
The stopping power, or stopping force, describes the energy loss per unit track length for a particle travelling through a medium. It is defined by
\begin{equation}\label{StoppingPowerMathematicalDefinition}
	-\frac{\D E}{\D x} = n\int_0^E E' \frac{\D \sigma}{\D E'} \D E',
\end{equation}
with $\D\sigma/\D E$ the differential collision cross section for a single particle collision, $n$ the number density of the medium, $E'$ the energy lost in a collision, and $E$ the kinetic energy of the decay product.

The stopping due to bound electrons can further be separated into two cases: projectile energy losses due to ionization and excitation. For the projectile velocities relevant here, the perturbation induced by the passing ion varies too rapidly for bound electrons to respond adiabatically. Electronic excitation, which requires a well‑defined energy transfer matching a discrete transition, becomes unlikely because the interaction time is shorter than the characteristic timescales of bound‑state dynamics \textcolor{black}{\citep{Darwin1912}}. In contrast, ionization remains probable because the continuum of available final states does not impose such resonance conditions. \textcolor{black}{Therefore, at these energies the the contribution to the stopping power from bound electrons is completely dominated by ionization compared to excitation.}

The differential cross section for such ionization events, derived under the assumption that the target mass is much smaller than the projectile mass, was formulated by \cite{Gryzinsky1965, Gryzinsky1965II, Gryzinsky1965III}. The derivation is fully classical\textcolor{black}{, as the initial velocities for alpha and fission particles are roughly $0.05c$, with $c$ the speed of light, making a relativistic treatment unnecessary, unlike for the much faster beta electrons.} Furthermore, Gryziński treats the interaction as a binary collision between the projectile and a bound electron, with the binding energy entering only as a threshold condition on the energy transfer. The projectile is assumed to follow a straight-line trajectory, and the target electron is taken to be initially at rest apart from its binding energy. Under these assumptions, the energy transfer is obtained from classical scattering kinematics, leading to an analytic expression for the differential ionization cross section that depends only on the projectile velocity and the electron binding energy. The stopping power is given by
\begin{equation}\label{boundstopping}
    -\frac{\D E}{\D x} = n_i \sum_{n}N_n\frac{\pi\left(Z_n e^2\right)^2}{I_n} G_S(v_p/v_n),
\end{equation}
where $n_i$ denotes the ion density in the plasma, the index $n$ sums over all electron orbitals of the target, $N_n$ is the electron occupation number of said orbitals, $Z_n$ \textcolor{black}{is the effective charge of the projectile as experienced by an electron in orbital $n$ (see section \ref{subsec:ionizationcrosssection}, Eq. \ref{ZnEq})}, $e$ is the elementary charge, $I_n$ the electron binding energy in orbital $n$, $v_p$ the projectile velocity and $v_n = \sqrt{2I_n/m_e}$ the velocity of the bound electron with $m_e$ the electron mass. The velocity matching function $G_S$ is defined by

\begin{equation}
G_S(V) =
\begin{cases}
f_1(V)\,f_2(V)\,f_3(V), & V > 0.243,\\[6pt]
\displaystyle \frac{128}{45}V^4, & V \le 0.243,
\end{cases}
\label{eq:GS}
\end{equation}

\noindent\textit{with}
\begin{align}
f_1(V) &= \frac{V}{(1+V^2)^{3/2}},\nonumber\\
f_2(V) &= \frac{m_e/m_p}{1+V^2}\log(\alpha_S)
      + \frac{4}{3}\left(1-\frac{1}{\alpha_S}\right)\log(2.7 + V),\nonumber\\
f_3(V) &= 1 - \alpha_S^{-(1+V^2)}, \nonumber
\end{align}
\textcolor{black}{where $m_p$ is the projectile mass,} $\alpha_S = 4V(1+V)$ and the value of $V\approx 0.243$ makes the piecewise function continuous. \textcolor{black}{The function $G_S$ accounts for optimal energy loss, peaking around $V\approx1.6$.}

To obtain consistent and shell-resolved electron binding energies for the
ionization model, atomic structure calculations are performed using the
\textsc{PySCF} package~\citep{PySCF2018}. All calculations employed the
relativistic X2C Hamiltonian together with the all-electron
\texttt{x2c-QZVPPall} basis set, which provides high-quality orbital energies suitable for use as binding energies in the stopping-power calculations.
Using these binding energies, the stopping powers calculated for a neutral xenon target gas agree with the tabulated values in the NIST ASTAR database \citep{Berger2005} within the energy interval relevant for thermalization.

For the contribution to the projectile stopping due to free plasma electrons, another approach is taken. Since the electron temperature in the nebular phase of the merger ejecta has dropped significantly, the velocity of the fission and alpha fragments far exceeds that of thermal electrons. In this regime, the stopping power due to thermal electrons is given by the 
Bohr formula \citep{Borh1912}:
\begin{equation}\label{BohrThermal}
    -\frac{\D E}{\D x}
    = \left(\frac{Z_{\mathrm{eff}} e \,\omega_p}{v_p}\right)^2
      \ln\!\left(\frac{m_e v_p^3}{Z_{\mathrm{eff}} e^2 \omega_p}\right),
\end{equation}
where \(Z_{\mathrm{eff}}\) is the effective projectile charge, \(e\) the 
elementary charge, \(v_p\) the projectile velocity, \(m_e\) the electron mass, 
and \(\omega_p\) the electron plasma frequency 
\(\omega_p = \sqrt{4\pi n_e e^2 / m_e}\), with \(n_e\) the free–electron density.
The total stopping power is then obtained by summing the contributions from 
bound and free electrons.

\begin{figure*}
    \centering
    \begin{subfigure}[t]{0.49\textwidth}
        \centering
        \includegraphics[width=\textwidth]{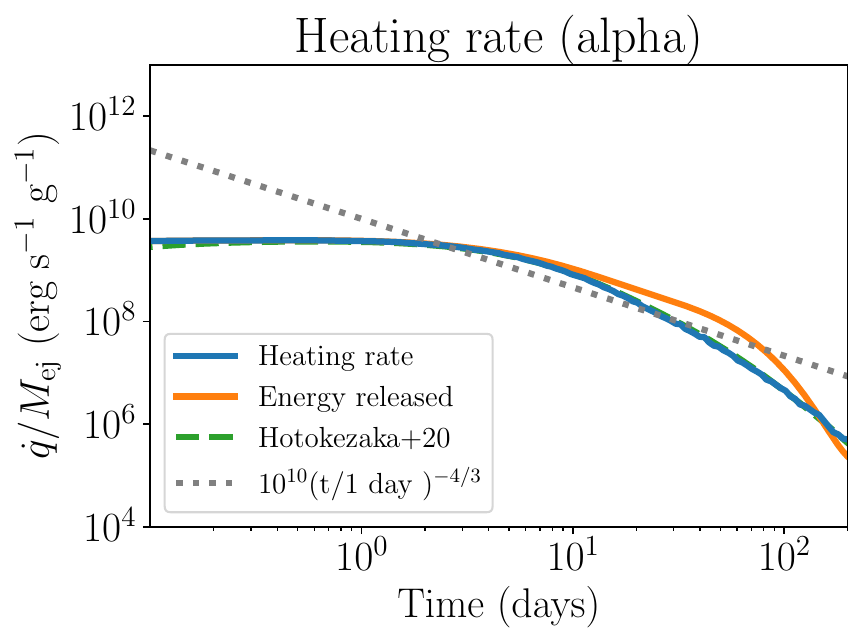}
        \label{fig:AlphaHeating}
    \end{subfigure}
    \hfill
    \begin{subfigure}[t]{0.49\textwidth}
        \centering
        \includegraphics[width=\textwidth]{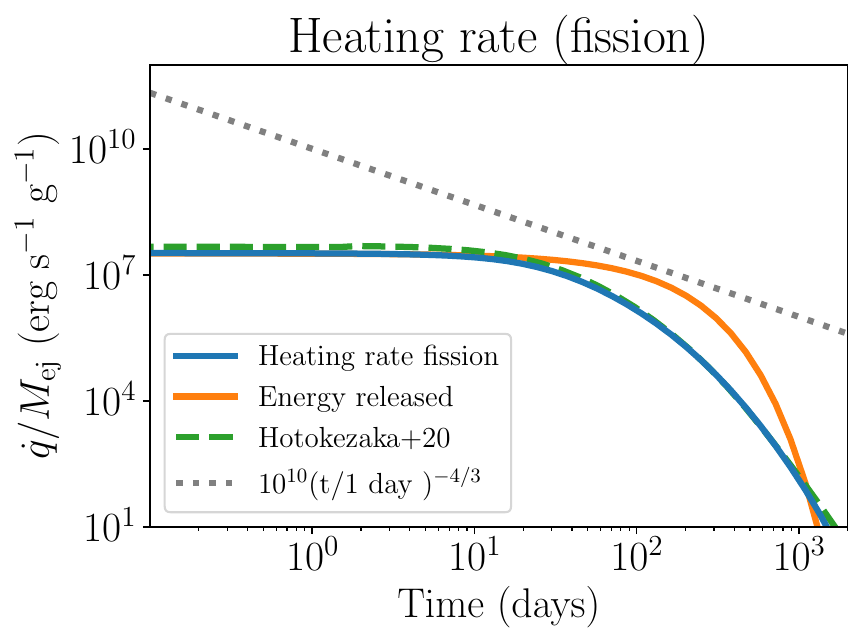}
        \label{fig:FissionHeating}
    \end{subfigure}
    \caption{Plots of the heating rate (blue) found for fission fragments and alpha decay particles \textcolor{black}{in a Xe II medium,} normalized by the ejecta mass $M_{\rm ej}=0.05 M_\odot$. This is compared to the respective \textcolor{black}{total kinetic energy released during the radioactive decay of the species} per unit time (orange). Initial abundances for alpha decay are $Y$(Rn-222) $=4.0\times 10^{-5}$, $Y$(Ra-223) $=2.7\times 10^{-5}$, $Y$(Ra-224) $=4.1\times 10^{-5}$ and $Y$(Ra-225) $=2.7\times 10^{-5}$. For spontaneous fission an abundance $Y$(Cf-254) $=2.0\times 10^{-6}$ has been placed. All abundances \textcolor{black}{for the parent nuclei} are taken from  \protect\cite{Wu2019}. The dashed gray line represents the analytical heating rate by beta decay, \textcolor{black}{and the dashed green line represents the heating rates found in \protect\cite{Hotokezaka2020}.}}
    \label{fig:combinedHeating}
\end{figure*}

\subsection{Effective projectile charge} \label{subsec:effectiveCharge}
\textcolor{black}{The remaining quantity to be determined in the stopping power for alpha particles and fission fragments is the effective projectile charge, \textcolor{black}{$Z_\text{eff}$}, introduced in the previous section.} Particular care must be taken when analysing the thermalization of fission fragments, as they retain bound electrons, whereas $\alpha$-particles remain fully ionized. At velocities relevant to this context, charge transfer is the primary mechanism for increasing the number of bound electrons, while single ionization serves as the dominant process for their reduction. Using the charge change rates obtained in \cite{Peter1991}, it becomes clear that \textcolor{black}{the rate at which a single particle thermalizes in the expanding medium, defined as $\alpha_{\rm th} = \D \ln(E)/\D t$ (see Eq. \ref{energyLoss}) is approximately four orders of magnitude smaller than the capture and ionization rate.} \textcolor{black}{As a consequence, the effective charge of the fission fragment is assumed to evolve only through the effective charge equilibrium values corresponding to solving the previously described rate equations for a fission fragment travelling at constant velocity through the medium.} These equilibria are well approximated with an empirical formula found by \cite{Schiwietz2001}. The fit-formula is given as

\begin{equation}\label{SchiwietzFitFormula}
	Z_{\rm eff} = Z_p \frac{376x + x^6}{1428-1206x^{0.5} + 690x},
\end{equation}
with
\begin{equation}
	x = \left(\frac{v_p}{v_B}Z_p^{-0.52}Z_t^{0.03-0.017 Z_p^{-0.52}v_p/v_B}\right)^{1+0.4/Z_p},
\end{equation}
with $Z_p$ the proton number of the projectile atom $Z_t$ the proton number of the target atoms, \textcolor{black}{the Bohr velocity $v_B = 2.19\times10^8\,{\rm cm\,s^{-1}}$ and again $v_p$ the velocity of the projectile particle}.

\subsection{Expanding Medium}\label{subsec:expandingmedium}
In addition to the loss of energy via collisions, the particles travel in an expanding medium and therefore lose energy via adiabatic expansion. The energy loss rate for a sub-relativistic projectile is given by
\begin{equation}\label{energyLoss}
	\frac{\D E}{\D t} = - \frac{3(\gamma-1)E}{t} - \left(v_p\frac{\D E}{\D x}\right)_{\rm plasma},
\end{equation}
\textcolor{black}{where the first term on the right-hand side represents the projectile energy loss due to the adiabatic expansion of the medium, with adiabatic index $\gamma = 5/3$ for non-relativistic particles like the ones we discuss. The second term in Eq. \ref{energyLoss} represents the energy loss rate due to plasma collisions, summing the free and bound contributions Eq. \ref{boundstopping} and Eq. \ref{BohrThermal} respectively and multiplying by $v_p$ to make it an energy loss rate. By introduction of the stopping cross $K_{st}(t,v_p)$ section we may write}
\begin{equation}\label{energyLossRate}
	\left(\frac{\D E}{\D t}\right)_{\rm plasma} = \left(v_p\frac{\D E}{\D x}\right)_{\rm plasma} = -v_p\rho_m(t)K_{\rm st}(t,v_p),
\end{equation}
where $K_{\rm st}(t,v) = K_{\rm st, free}(t,v_p) + K_{\rm st, bound}(t,v_p)$ and a mass weight density $\rho_m = C_\rho M_{\mathrm{ej}} t^{-3}v_0^{-3} = \rho_0 t^{-3}$, with $C_\rho$ a normalization constant under the assumption of a power law mass density profile \citep{Hotokezaka2020}. \textcolor{black}{The stopping powers resulting from Eq. \ref{energyLossRate} are shown in Fig. \ref{fig:combined}. It is clear that in both the fission fragment and alpha particle cases the stopping powers for collisions with bound and free electrons are similar around the initial energies of the particles (around 7 MeV and 130 MeV for alpha and fission respectively), and stopping due to collisions with thermal electrons start to dominate as kinetic energy decreases.}

As shown in Fig. \ref{fig:StoppingCrossSectionFissionAlpha}, \textcolor{black}{the stopping cross sections rescaled with velocity $\beta = v_p/c$ with $c$ the speed of light follow a power law in energy for velocities relevant for thermalization.} Because of this, Eq. \ref{energyLossRate}  is well approximated by the following
\begin{equation}\label{energyLossDifferentialEquation}
	\left(\frac{\D E}{\D t}\right)_{\rm plasma} = -\rho_m(t)\left[v_p K_{\rm st}(t,v_p)\Big|_{t = t_0}\right] \left( \frac{E}{E_0}\right)^{a},
\end{equation}
where $t_0$ is the time at which a fission or alpha decay fragment is released with an initial energy $E_0$ and defining $a$ as the power-law index. \textcolor{black}{Now, Eq. \ref{energyLossDifferentialEquation} can now be substituted back into Eq. \ref{energyLoss} that includes the adiabatic loss, which in our conditions is given by $-2E/t$.} This results in a Bernoulli type differential equation and has an analytical solution
\begin{equation}\label{EnergyLossIncludingAdiabatic}
\begin{aligned}
&E(t,t_0,E_0) =   \\
&\Bigg(E_0^{1-a}\left(\frac{t_0}{t}\right)^{2(1-a)}+ \frac{(a-1)\,\rho_0\,(v_pK_{\rm st})_0\,E_0^{-a}
\left[\left(\frac{t}{t_0}\right)^{2a}-1\right]}{2a\,t^2}
\Bigg)^{\!1/(1-a)} ,
\end{aligned}
\end{equation}
where $a$ is the power-law index from Eq. \ref{energyLossDifferentialEquation}.
By means of Eq. \ref{EnergyLossIncludingAdiabatic} the kilonova heating rate at a time $t$, as done by a single decay product emitted at a time $t_0$ is given by
\begin{equation}
\begin{aligned}
\dot{q}_i(t,t_0)
&= \rho_m(t)\, v_p\, K_{\rm st}\!\left(t, E(t,t_0,E_{0,i})\right) \\
&= \rho_m(t)\,
   \sqrt{\frac{2E(t,t_0,E_{0,i})}{m_p}}\,
   K_{\rm st}\!\left(t, E(t,t_0,E_{0,i})\right).
\end{aligned}
\end{equation}
Approximating the radioactive decay rate of a nucleus as $\dot{N}_i \approx N_i/\tau_i$, where $\tau_i$ is the nucleus lifetime. For alpha decay in particular, decay chains have to be taken into account, here only dominant decay modes are considered and obtained from \cite{IAEA_LiveChart}. The evolution of abundances are given by the Bateman equations \citep{Bateman1910}. The total heating rate at a time $t$ is then given by \citep{Hotokezaka2020}
\begin{equation}
	\dot{q}(t) = \sum_i \int_{t_{0, i}}^t \D t_0 \hspace{2px} \dot{q}_i(t,t_0)\frac{N_i(t_0)}{\tau_i},
\end{equation}
where $t_{0, i}$ is the time at which particles that are  thermalized at time $t$ are synthesized. Figure \ref{fig:combinedHeating} shows the heating rates \textcolor{black}{$\dot{q}/M_{\rm ej}$} of fission and alpha decay, which is in accordance with \cite{Hotokezaka2020} despite \textcolor{black}{the change of stopping cross sections $K_{\rm st}$ in Eq. \ref{energyLoss}. The slight difference in heating rate for the fission fragments is attributed to a small difference in the chosen injection energies.}.

\section{Ionization for Alpha and Fission} \label{sec:ionization}
Ionization processes in plasmas are governed by the interaction between energetic particles and bound electrons. The probability of ionization is quantified by the ionization cross section, which depends on the projectile's effective charge and velocity relative to the electron's orbital motion. To translate this microscopic interaction into macroscopic ionization rates, one must consider the particle energy spectrum, which describes how charged particles lose energy as they traverse a medium. This spectrum is shaped by both collisional losses and the injection profile of the particles, and is formally described by the Spencer–Fano equation.

\subsection{Ionization Cross Section and Instantaneous Spectrum} \label{subsec:ionizationcrosssection}

The ionization cross section for an electron in the $n$th shell follows
directly from the differential ionization cross section introduced earlier in the context of the stopping power. Integrating the differential expression over all allowed energy transfers gives
\begin{equation}\label{ionizationcrosssection}
    \sigma_{\mathrm{ion},n}(v_p)
    = \int_{0}^{E(v_p)}
      \frac{\mathrm{d}\sigma_n}{\mathrm{d}E'}\,\mathrm{d}E'
    = \pi \left( \frac{Z_n e^{2}}{I_n} \right)^{2}
      G\left( \frac{v_p}{v_n} \right),
\end{equation}
where $Z_n$ denotes the effective charge experienced by the electron in orbital $n$, $e$ is the elementary charge, $I_n$ the corresponding binding energy, and $G$ the corrected velocity-matching function that relates the projectile velocity $v_p$ to the characteristic orbital velocity $v_n$, \textcolor{black}{peaking around $v_p \approx v_n$ for optimal ionization\footnote{Though resembling both the form and role of the earlier defined $G_S$ (see Eq. \ref{boundstopping}), their difference originates from the definition of the stopping power (Eq. \ref{StoppingPowerMathematicalDefinition}) and ionization cross sections (the first equality in Eq. \ref{ionizationcrosssection}). Detailed calculations are presented in \cite{Gryzinsky1965II}.}}. This provides a self-consistent description of the stopping and ionization process needed to guarantee the accuracy of the ionization efficiency. The corrected velocity matching function is defined as (\citealt{McGuire1973})
\begin{equation}
G(V) =
\begin{cases}
g_1(V)\,g_2(V)\,g_3(V), & V > 0.215,\\[6pt]
\displaystyle \frac{4}{15}V^4, & V \le 0.215,
\end{cases}
\label{eq:G}
\end{equation}

\noindent\textit{with}
\begin{align}
g_1(V) &= \frac{\alpha^{3/2}}{V^2},\\
g_2(V) &= (1-\beta)\left(1-\beta^{1+V^2}\right),\\
g_3(V) &= \alpha + \frac{2}{3}(1+\beta)\ln(2.7 + V),
\end{align}
where the parameters $\alpha$ and $\beta$ are defined as
\begin{equation}
	\alpha = \frac{V^2}{1+V^2}, \quad \beta = \frac{1}{4V(1+V)}.
\end{equation}

The screened or effective projectile charge as experienced for an electron in orbital $n$ that is presented in Eqs. \ref{boundstopping} and \ref{ionizationcrosssection} can be constructed by solving for the number of electrons inside a spherical volume of radius $b_n = \sqrt{\sigma_{{\rm ion},n}/\pi}$. \textcolor{black}{Another representation of $Z_n$ can be obtained by rewriting Eq. \ref{ionizationcrosssection}. These two representations give a system of equations that now allow us to solve for $Z_n$ and $b_n$, as presented in \cite{Peter1991II}:}
\begin{equation}\label{ZnEq}
\begin{cases}
\displaystyle
Z_n = Z_p - \sum_{n'} N_{n'} 
\int_{0}^{b_n} \mathrm{d}^3\mathbf{r}\, \left|\psi_{n'}(\mathbf{r})\right|^2,
\\[1.2em]
\displaystyle
Z_n = \frac{I_n\, b_n}{e^{2}\sqrt{G\!\left(v_{p}/v_n\right)}}.
\end{cases}
\end{equation}
The normalized wave function of each orbital $\psi_n$ is obtained from the same \textsc{PySCF} calculations used to determine the binding energies, guaranteeing internal consistency between the orbital shapes and their corresponding energies.

The construction of the ionization rate from the particle degradation spectrum has been explored in several early studies, including the influential treatment by \cite{Axelrod1980} in the context of radioactive energy deposition in supernovae ejecta. In the case of heavy and sub-relativistic particles, the energy loss per collision compared to the kinetic energy is substantially reduced in comparison to beta electrons. The energy spectrum of a charged particle slowing down in a medium is determined by the balance between collisional energy losses and the injection of new particles from a source. \cite{Spencer1954} derived an integral equation that captures this balance in steady state, now known as the \emph{integral form of the Spencer-Fano equation}:  

\begin{equation}\label{SpencerFano}
    y(E)\int_0^E \D \epsilon \, k(E, \epsilon) 
    = \int_0^\infty \D \epsilon \, y(E + \epsilon)k(E + \epsilon, \epsilon) + s(E),
\end{equation}
where $s(E)$ denotes the source term, representing a uniformly distributed number of particles injected per unit volume, time, and energy (cm$^{-3}$ s$^{-1}$ erg$^{-1}$). The function $k(E,\epsilon)\D \epsilon$ gives the probability per unit path length (cm$^{-1}$) that a particle of energy $E$ loses an amount $\epsilon$. Finally, $y(E)$ is the particle energy spectrum, expressed as flux per unit energy (cm$^{-2}$ s$^{-1}$ erg$^{-1}$). The stopping power is then defined as  

\begin{equation}
    S(E) = \int_0^E \D \epsilon \, \epsilon \, k(E, \epsilon).
\end{equation}

For a mono-energetic source, the function takes the form $s(E) = s_0 \delta(E - E_0)$, where $s_0$ is the source strength (cm$^{-3}$ s$^{-1}$). Introducing a normalized spectral function $\bar{y}(E) \equiv y(E)/s_0$ with units cm erg$^{-1}$, the Spencer-Fano equation becomes  
\begin{equation}\label{SF2}
    \bar{y}(E)\int_0^E \D \epsilon \, k(E, \epsilon) 
    = \int_0^\infty \D \epsilon \, \bar{y}(E + \epsilon)k(E + \epsilon, \epsilon) + \delta(E - E_0).
\end{equation}
The maximum energy transfer in a collision with an electron is $2 m_e v_p^2$, which is much smaller than the kinetic energy of a fission fragment or $\alpha$-particle, scaling as $m_e/m_p \ll 1$. Consequently, $k(E,\epsilon)$ has support only for $\epsilon \ll E$. Expanding Eq.~\ref{SF2} under this assumption, and introducing a cut-off $\epsilon \ll \epsilon' \ll E$, yields  
\begin{equation}
\begin{aligned}
\bar{y}(E)\!\int_0^{\epsilon'} \! \mathrm{d}\epsilon\, k(E,\epsilon)
\approx \int_0^{\epsilon'} \! \mathrm{d}\epsilon\, \Big[
& \bar{y}(E)\,k(E,\epsilon)  \\
& +\, \bar{y}(E)\,\pdv{k(E,\epsilon)}{E}\,\epsilon  \\
& +\, k(E,\epsilon)\,\pdv{\bar{y}(E)}{E}\,\epsilon
\Big]
+ \delta(E - E_0).
\end{aligned}
\end{equation}
For $E \neq E_0$, the solution is simply
\begin{equation}\label{instantSpectrum}
    \bar{y}(E) \approx \left(\int_0^{\epsilon'} \D \epsilon \, \epsilon \, k(E, \epsilon)\right)^{-1} = \frac{1}{S(E)}.
\end{equation}

Having established the form of the energy spectrum, the number of ionizations $\mathcal{N}_i$ can be determined directly. Since $\bar{y}(E)$ describes the effective path length of a particle with energy $E$, and given the ion number density $n_{\text{ion}}$ and ionization cross section $\sigma_{\text{ion}}(E) = \sum_n \sigma_{\text{ion},n}(E)$, the ionization yield is  

\begin{figure*}
    \centering

    \begin{subfigure}{0.49\textwidth}
        \centering
        \includegraphics[width=\textwidth]{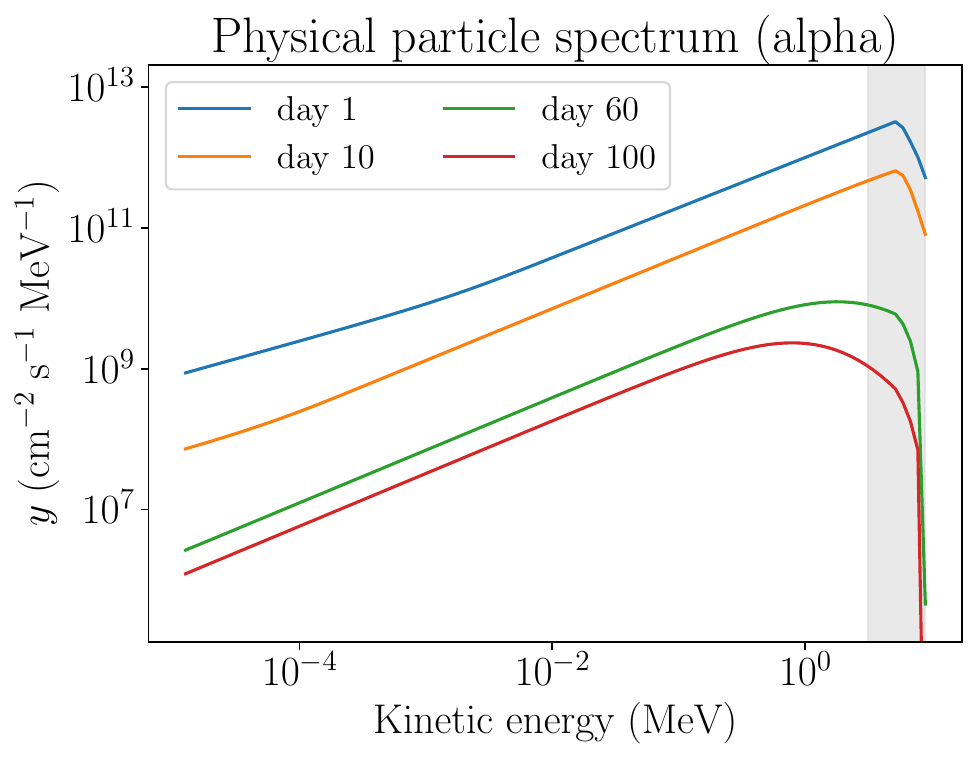}
        \label{fig:SpecFission}
    \end{subfigure}
    \hfill
    \begin{subfigure}{0.485\textwidth}
        \centering
        \includegraphics[width=\textwidth]{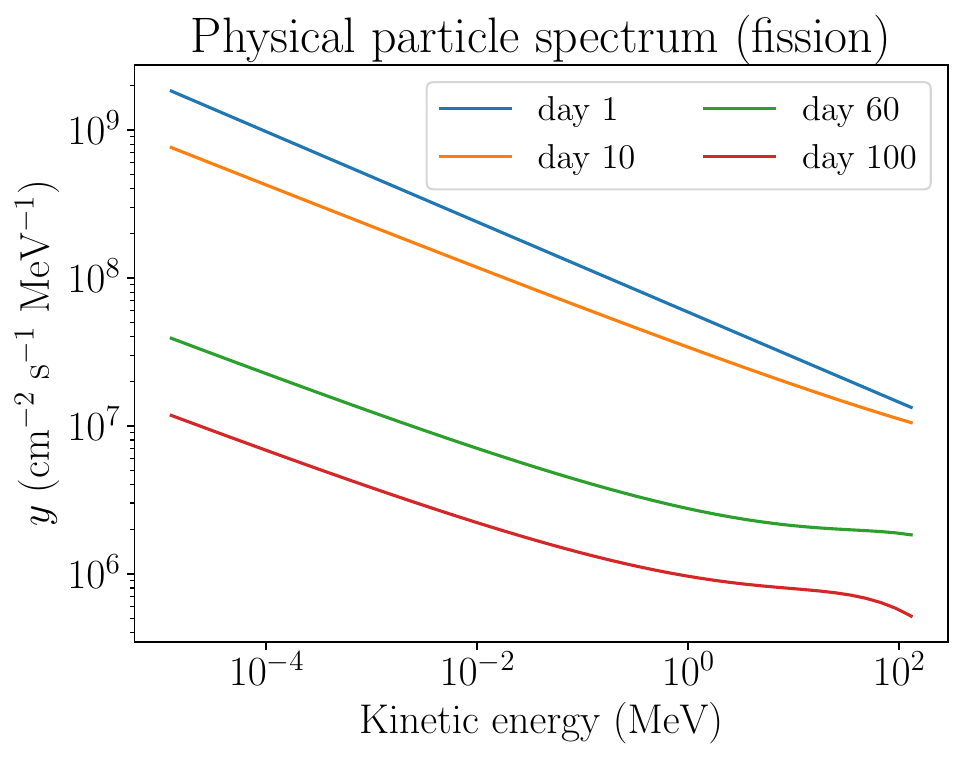}
        \label{fig:SpecAlpha}
    \end{subfigure}

    \caption{Particle spectra for fission and $\alpha$-particles coming from the same ejecta \textcolor{black}{and parent nuclei} configuration as Fig. \ref{fig:combinedHeating}. For the alpha case, the range of injection energies is represented by the gray area, with larger decay energies being many times more numerous at earlier times. \textcolor{black}{Note the characteristic presence of the result $y(E)\propto 1/S(E)$ when looking at early times or low energies (low energies are synonymous with particles emitted at earlier times). This should be compared with Fig. \ref{fig:combined}. }}
    \label{fig:ParticleSpectra}
\end{figure*}

\begin{equation}\label{numberOfIonizations}
    \mathcal{N}_i = \int_{E_{\text{th}}}^{E_{0,i}} \D E \, n_{\text{ion}} \, \sigma_{\text{ion}}(E) \, \bar{y}(E),
\end{equation}
where $E_{\text{th}}$ is the \textcolor{black}{kinetic energy of the projectile at which it is considered a part of the thermal bath.}  

It should be noted that this expression accounts only for primary ionizations. In reality, additional ionizations occur due to secondary electrons liberated in the process. For fission fragments, these secondary electrons have kinetic energies in the range $E_{\min} \approx 250 \, \text{eV} < E_s < E_{\max} \approx 200 \, \text{keV}$, where $E_{\min}$ is set by the minimum Coulomb energy transfer and $E_{\max}$ by the kinematic limit, and strongly dominated by the lower energy tail. In the present work, we therefore safely neglect any contributions from these secondary electrons for the ionizing fission fragments and $\alpha$-particles.

\subsection{Physical Particle Spectrum}
In the previous case, the degradation spectrum for fission and $\alpha$-particles describes the steady-state number of particles that travel through the medium. There is no notion of adiabatic expansion. Since thermalization is described via the continuous slowing down approximation, the construction of a continuity equation is permitted. The goal is to obtain a single-particle spectrum, meaning the energy evolution of a particle is only time dependent. The continuity equation is given by
\begin{equation}\label{continuityEquation}
    \dv{N}{t} + N\pdv{\dot{E}}{E} = Q,
\end{equation}
where the particle spectrum $N(E,t)$ denotes the number of particles per unit energy, and the source term $Q(E,t)$ denotes the number of particles produced by decay per unit energy per unit time (for exponential decay this is just $\exp(-t/\tau)/\tau$). Since the decays discussed are only mono-energetic, the source term is given by $Q(E,t) = \dot{N}_{\rm inj}\delta(E(t)-E_0)$, where $\dot{N}_{\rm inj}$ is the number of particles that are released by a decay at time $t_0$. Since Eq. \ref{continuityEquation} is a linear first order ordinary differential equation, the solution is given by
\begin{multline}\label{ParticleSpectrum}
    N(E,t) = \frac{\dot{N}_{\rm inj}(t_0)}{\|\dot{E}(E_0,t_0)\|}\exp{-\int_{t_0}^{t}\left(\pdv{\dot{E}}{E}\right)\mathrm{d}t'}=  \frac{\dot{N}_{\rm inj}(t_0)}{\|\dot{E}(E_0,t_0)\|} \\\times\exp{\int_{t_0}^{t}\left(\frac{2}{t'} +a\rho_m(t')\left[v_p K_{\rm st}(t,v)\Big|_{t = t_0}\right] \left( E(t')\right)^{a-1} E_0^{-a}\right)\mathrm{d}t'}.
\end{multline}

\textcolor{black}{Using the power-law approximation for the stopping power (Eq.~\ref{energyLossDifferentialEquation}), which remains valid over several orders of magnitude in energy below the initial kinetic energy of the decay particles (see Fig.~\ref{fig:StoppingCrossSectionFissionAlpha}), and including adiabatic expansion losses, the energy evolution is described by Eq.~\ref{EnergyLossIncludingAdiabatic}.} Finally, because the particles move along a trajectory and strictly decrease in energy, there is a one-to-one correspondence to be made between a particles current energy $E(t)$ and the time $t_0$ when it entered the medium. This relation must be numerically inverted to obtain $N(E,t)$ instead of $N(t_0, t)$. The presented integral has to be performed numerically.

The physical spectra for fission and $\alpha$-particles are presented in Fig. \ref{fig:ParticleSpectra}, where the spectrum $y(E,t) = n(t)v_p(E)N(E,t)$, with $v_p$ being the speed of the fission or $\alpha$-particles and $n$ denotes the number density of the parent nucleus responsible for this decay in the ejecta. The spectra exhibit clear transitions between physical regimes. At early times, the ejecta density is high and collisional energy loss dominates. Under these conditions, decay products thermalize on timescales much shorter than the dynamical time of the ejecta, and the spectra reduce to the instantaneous form of Eq.~\ref{instantSpectrum} \textcolor{black}{(compare this with the shape of the stopping powers in Fig. \ref{fig:combined}).} At later times, however, the ejecta has expanded and its density has dropped substantially. Adiabatic losses then become the dominant energy-loss channel, and decay products require much longer to thermalize.

This behaviour imprints itself on the spectra. The low-energy portion (below $\sim 0.1$\,MeV) is populated mainly by particles produced around $\sim 20$ days after merger, when collisional losses still shaped the spectrum; adiabatic expansion alone introduces no structure because it scales linearly with kinetic energy. At higher energies, the spectrum flattens at late times due to the dominance of adiabatic cooling. This flattening is partially masked by the exponential decay of the parent nuclei, which suppresses the high-energy end of the spectrum at late times.

\begin{table*}
\centering

\begin{subtable}[t]{0.2\textwidth}
\centering
\begin{tabular}{lcc}
\hline
\textbf{Target} & \textbf{WIP} \\
\hline
Xe\,I   & 2.6/2.5  \\
Xe\,II  & 6.7/6.5  \\
Xe\,III & 8.8/8.5  \\
Xe\,IV  & 14.3/13.6 \\
Xe\,V   & 18.1/16.9 \\
\hline
\end{tabular}
\caption{Xenon charge states}
\label{XenonSubfig}
\end{subtable}
\hfill
\begin{subtable}[t]{0.7\textwidth}
\centering
\begin{tabular}{lccccc}
\hline
\textbf{Target/Medium} & He\,I & He\,II & Sr\,II & Xe\,II & Nd\,II \\
\hline
He\,I  & 2.3/2.3 & 11.7/11.7  & 20.6/20.6 & 23.1/22.7 & 24.8/24.4 \\
He\,II & 13.0/12.6 & 30.1/28.9 & 53.9/51.8 & 60.7/57.4 & 65.2/61.6 \\
Sr\,II & 2.4/2.4 & 8.0/8.1 & 14.2/14.3 & 16.0/15.7 & 17.1/16.9 \\
Xe\,II & 0.9/0.9 & 3.6/3.3 & 6.0/5.9 & 6.7/6.5 & 7.2/7.0 \\
Nd\,II & 1.2/1.2& 4.1/4.2 & 7.2/7.3 & 8.1/ 8.1& 8.9/8.7\\
\hline
\end{tabular}
\caption{Work per ion pair for various media}
\label{VariousSubfig}
\end{subtable}
\caption{Work per ion pair for fission fragments and alpha‑particles (fission WIP/alpha WIP).
Table \ref{XenonSubfig} shows results for neutral ejecta composed solely of xenon in different charge states, and thermal electrons with abundance fractions $\chi_e = {0,1,2,3,4}$ from top to bottom.
Table \ref{VariousSubfig} shows results for various media, where particle spectra are generated using collisions with the listed ions (columns) but ionization cross sections are substituted from other species (rows). We adopt $\chi_e = 0.1$ for He I and $\chi_e = 1$ otherwise. All values are in units of $I$.}
\label{WorkPerIons}
\end{table*}

\subsection{Ionization Efficiency}\label{subsec:alphafissionionization}

With the results from the previous two sections, a measure for the ionization efficiency can be computed, with the work per ion pair being defined as 
\begin{equation}
    w_i=\frac{E_{0,i}}{\mathcal{N}_i}.
\end{equation}
For fission fragments and $\alpha$-particles, we find values for the work per ion pair between $w = 2.5I$ and $w = 18.1I$ for Xe I and Xe V respectively (see Table \ref{XenonSubfig}), where $I$ is the ionization potential of the most loosely bound electron. Here, the ejecta are assumed to be quasi-neutral and consist only of thermal electrons and one of the xenon-ion species. We note that the work per ion value for alpha decay particles in a Xe I gas is consistent with experiments recorded in the literature \citep{Giesen2014, Baek1995-zz}. 

In Table \ref{VariousSubfig}, we show work per ion pair for cases in which the particle spectra of fission fragments and $\alpha$-particles are fixed by the indicated medium, while the ionization cross sections are substituted with those of other ions. \textcolor{black}{More explicitly, the abundances of the parent nuclei that produce the fission and alpha particles are the same as those in Fig. \ref{fig:combinedHeating}.} The various media are He I with an electron abundance fraction $\chi_e = 0.1$, and for He II, Sr II, Xe II and Nd II we fixed $\chi_e = 1$. \textcolor{black}{The choice of the medium sets the particle spectrum by solving Eq. \ref{ParticleSpectrum}, whereas the choice in target sets the ionization cross section $\sigma_{ion}$ used only in \ref{numberOfIonizations} and has no influence on the choice of $\bar{y}$.} The results again show no significant difference in ionization efficiency between $\alpha$-particles and fission fragments. Both projectiles follow the same underlying physics and are far more massive than the bound and free electrons they ionize, leading to nearly identical work per ion pair. In general, $\alpha$-particles appear slightly more efficient, a consequence of treating their charge as a constant $+2$, whereas fission fragments undergo charge evolution and a slight increase in typical initial kinetic energy over fission fragments.

\section{Ionization and spectrum for $\beta$-
electrons}\label{BetaIonSection}

\begin{figure*}
    \centering

    \begin{subfigure}[t]{0.5\textwidth}
        \vspace{-179pt}
        \centering
        \includegraphics[width=0.99\textwidth]{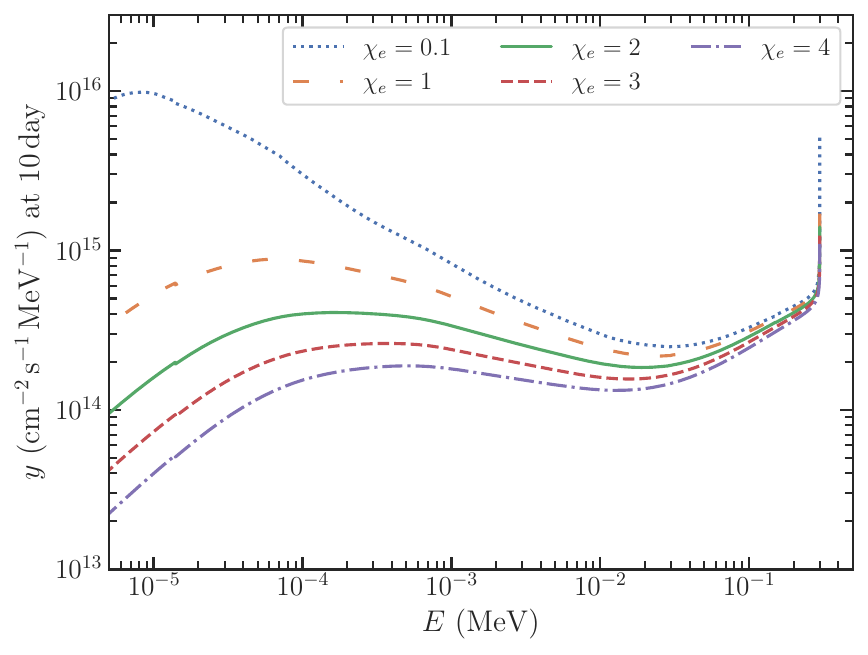}
        \label{fig:Specbeta}
    \end{subfigure}
    \hfill
    \begin{subfigure}[t]{0.485\textwidth}
        \centering
        \includegraphics[width=\textwidth]{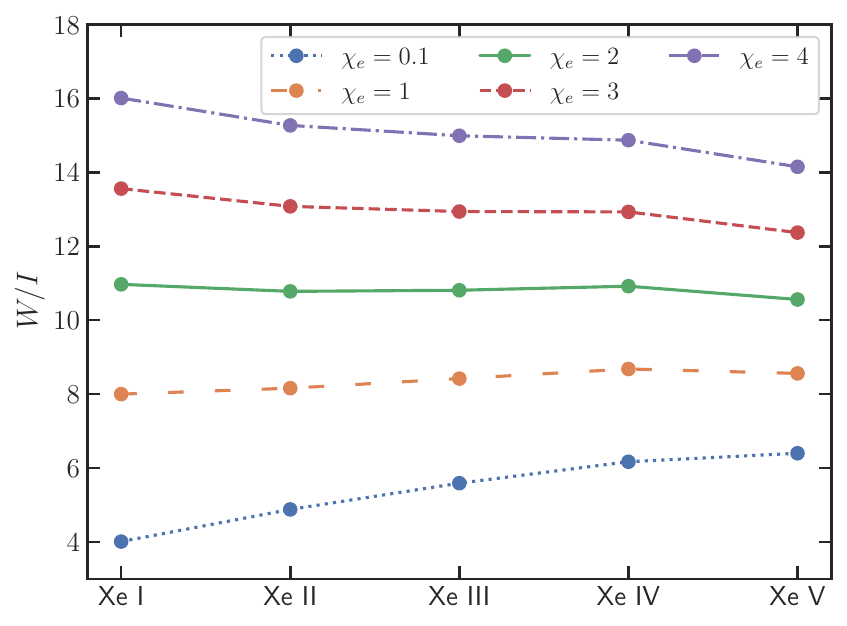}
        \label{fig:SWPI_beta}
    \end{subfigure}

    \caption{Electron spectra (\textit{left}) for the same ejecta configuration as Fig. \ref{fig:combinedHeating}. Work per ion pair for $\beta$-electrons and their secondaries for different ionization stages of xenon (\textit{right}).}
    \label{fig:ParticleSpectrabeta}
\end{figure*}

In contrast to heavy charged particles, the transport and thermalization of $\beta$-electrons in kilonova ejecta have been studied extensively \citep{Barnes2016,Waxman2019ApJ,Kasen2019ApJ,Hotokezaka2020}, and their role in powering the light curve is comparatively well understood. Most existing work, however, focuses on their heating efficiency rather than their detailed spectral evolution or ionization behaviour. To place our heavy-particle results in context, we therefore also examine the $\beta$-electron spectrum and its associated work per ion pair.

For the $\beta$-electron spectrum, the same principles apply as for the fission and $\alpha$-particles, though the degradation of the electrons occurs through a wider selection of channels. The degradation equation is 
\begin{align}
\bar{y}(E)\!\int_{\epsilon_0}^{E} k(E,\epsilon)\,d\epsilon
= {} &
\sum_n \bar{y}(E+\epsilon_n)\,
       k_n(E+\epsilon_n,\epsilon_n) \nonumber \\[4pt]
& + \sum_{n}\int_{I_{n}}^{\lambda_{n}}
      \bar{y}(E+\epsilon)\,
      k_{n}(E+\epsilon,\epsilon)\,d\epsilon \nonumber \\[4pt]
& + \sum_{n}\int_{2E+I_{n}}^{E_0}
      \bar{y}(E')\,
      k_{n}(E',E+I_{n})\,dE' \nonumber \\[4pt]
& + \sum_{n}\eta_{n}\,
      \delta(E-\epsilon_{n})
      \int_{I_{n}}^{E_0}
      \bar{y}(E')\,\sigma_{n}(E')\,dE' \nonumber \\[4pt]
& + \dv{E}\!\left[L_{\rm th}(E)\,\bar{y}(E)\right]
  + \delta(E-E_0).
\label{eq:deg}
\end{align}
where $\epsilon_n$ is excitation energy and $I_n$ is the ionization potential, $E_0$ the injection energy and $\lambda_{n}={\rm min}(E_0-E,E+I_n)$. The first and second terms on the right-hand side of equation (\ref{eq:deg}) correspond  to the probability for the entry of electrons with $E'>E$ to $(E,E+dE)$ by excitation and ionization, respectively. The third term describes the generation of secondaries in $(E,E+dE)$.
The fourth term accounts for the production of Auger electrons following inner shell ionization, where the ionization of the $n$th shell is followed by an Auger electron of energy $\epsilon_{n}$ is created with a probability $\eta_{n}$. The fifth term is the energy loss due to Coulomb collision with thermal electron \citep[see][ for $L_{\rm th}$]{Kozma1992ApJ}. 
In ionized medium $\chi_e\gtrsim 1$, 
the contributions of excitation and Auger electrons are less than $10\%$, and therefore, we neglect these in this work \textcolor{black}{(for a discussion, see appendix \ref{app1}).}
The ionization cross sections by electron impacts are computed by using \texttt{HULLAC} (\citealt{HULLAC}).
The energy distribution of secondary electrons is assumed as  \citep{Opal1971}
\begin{align}
    P(E_p,E_s) = \frac{1}{J{\rm arctan((E_p-I)/J)}}\frac{1}{1+(E_s/J)^{2}},\label{eq:ckf}
\end{align}
where $E_p$ and $E_s$ are the primary and secondary energy, and $I$ is the ionization potential of the shell, and we use $J=0.7I$ \citep{Kozma1992ApJ}.

Figure \ref{fig:ParticleSpectrabeta} shows the electron degradation spectra in pure Xe medium with an initial kinetic energy of beta-particle of $E_{\beta}=300\,{\rm keV}$. The spectra for $\chi_e=(0.1,1,2,3,4)$ correspond to the medium composed of Xe I, Xe II, Xe III, and Xe IV, respectively. For $E\gtrsim E_{\beta}/3$, the electron spectra are proportional to the reciprocal of the stopping powers same as the cases for heavy particle. In the lower energy regime, $E\lesssim E_{\beta}/3$, the contribution of secondary electrons significantly increases the flux. The flux at lower energies decreases with increasing the electron fraction because the stopping due to the Coulomb collision is more efficient for higher $\chi_e$.
The ionization due to secondary electrons increases the ionization rates by a factor of $2$--$3$.

Using the electron spectra, we compute the work per ion pair, $w$, for the same selection of media and targets as discussed in section \ref{subsec:alphafissionionization}. Figure \ref{fig:ParticleSpectrabeta} ({\it right}) shows $w/I$, where $I$ is the first ionization potential of the target ion. For Xe ions, the values $w/I$ for different ionization stages are roughly the same for a given $\chi_e$ and increase with increasing $\chi_e$.
These results are quantitatively similar to those obtained for alpha decay and fission (see Table \ref{XenonSubfig}). \textcolor{black}{Table \ref{WorkPerIons2} presents the work per ion pair for beta-electrons under the same circumstances as in Table \ref{VariousSubfig}.} We also note that, among heavy-ion species, the work per ion pair scales with the effective number of electrons in the target ion, as discussed in Appendix \ref{app2}. This scaling can be used to estimate approximate values of the work per ion pair for various ion species. 

\begin{table}
\centering
\begin{tabular}{lccccc}
\hline
\textbf{Target/Medium} & He\,I & He\,II & Sr\,II & Xe\,II & Nd\,II \\
\hline
He\,I & 2.4 & 4.8 & 15.3 & 18.0 & 24.5 \\
He\,II & 6.0 & 10.6 & 37.3 & 44.5 & 47.5 \\
Sr\,II  & 2.5 &5.7& 16.5  & 19.1 & 20.6 \\
Xe\,II  & 1.1 & 2.4 & 7.1  & 8.2 & 8.8 \\
Nd\,II & 1.3 & 3.6 & 9.3  & 10.6  &  11.4\\
\hline
\end{tabular}
\caption{Work per ion pair for beta decay for various target ions and media. The values presented are in units of $I$, the first ionization potential for the given ionization state. We adopt $\chi_e=0.1$ for He I and $\chi_e=1$ otherwise. }
\label{WorkPerIons2}
\end{table}

\section{Discussion} \label{sec:resultsanddiscussion}
In this section, we discuss our findings on the heating and ionization caused by fission fragments and alpha decay particles. These results will be compared to the ionization efficiencies and spectra found for $\beta$-electrons.

\textcolor{black}{
Figure \ref{fig:combinedHeating} shows that the late-time heating rate follows power-law behaviour. Specifically, the fission heating rate scales as \( \dot{q}_{\text{fis}} \propto t^{-2.3} \) for \( t \approx \tau_{\text{fis}}\) which relates to the half life as $\tau_{fis}=t_{1/2}/\log(2)\approx 87.3 \text{ days}$, transitions to \( \dot{q}_{\text{fis}} \propto t^{-5} \) for \( t \gg \tau_{\text{fis}} \). For alpha decay at later times we find \( \dot{q}_\alpha \propto t^{-2.8} \). This behaviour arises from the approximately constant decay rate for alpha particles, which happens in general when many decay chains contribute. As the material expands, its density evolves as \( \rho_m(t) \propto t^{-3} \), which would be the heating rate if time and energy dependence was removed from the stopping cross section. Thus, the inclusion of time and energy dependence in the stopping cross section modifies this to \( t^{-2.8} \). For fission, the decay is not supported by decay chains and decreases exponentially, the expansion therefore further steepens the decline to \( t^{-5} \). This heating‑rate prescription remains consistent with \cite{Hotokezaka2020}, even though we adopt a different stopping‑power formulation to ensure self‑consistent thermalization and ionization.
}

In all three presented cases for the work per ion pair values, as the ambient material becomes more highly ionized, the ionization efficiency of the projectile naturally decreases. When a quasi-neutral single species ejecta is assumed as above, each increase in ionization state for the plasma ions will increase the number of thermal electrons while also leaving fewer bound electrons available for further ionization. Higher charge states also have more tightly bound outer electrons, making additional ionization progressively harder, though the effect of this is rather small for heavier elements such as xenon. For Xe II, the bound‑ and free‑electron stopping powers at the injection energies of both fission fragments and $\alpha$-particles are already comparable, reducing the ionization efficiency by roughly a factor of two relative to Xe I. This demonstrates that most ionizations occur while the projectile is still near its initial, highest velocity. The similar values for the stopping cross sections of Xe II and the thermal electrons further suggest a scaling relation: although xenon ions in these charge states still retain many loosely bound electrons, the number of thermal electrons increases with each ionization stage, leading to the approximate scaling
$$
\text{WIP} \approx \text{WIP}_{\chi_e=1} \times \left(\frac{\chi_e + 1}{2}\right),
$$
where $\text{WIP}_{\chi_e=1}$ is the work per ion pair for the case where $\chi_e=1$.

As noted in Section \ref{BetaIonSection}, the ionization efficiencies of fission fragments and $\alpha$-particles closely resemble those of $\beta$-electrons (see Tables~\ref{VariousSubfig} and \ref{WorkPerIons2}). This similarity is most pronounced for heavier ionization targets such as Sr~II, Xe~II, and Nd~II, whose fuller and more complex electron shells provide many loosely bound outer electrons (see also Appendix \ref{app2}). In contrast, for He~I--II the ionization efficiency differs by a factor of a few. This arises from the larger binding energies and the limited number of outer-shell electrons available to fission fragments and $\alpha$-particles, an effect that is less significant for the $\beta$-electron ionization cross sections and therefore produces the observed discrepancy.

\textcolor{black}{Since this work neglects the contributions of electron excitations in the computation for the work per ion pair because of increased computational complexity, the work per ion pair values especially for heavy targets is likely to be slightly underestimated. For xenon, excitation contributes roughly 20\% of the total work per ion pair, as inferred from the energy required to produce each scintillation photon \citep{Henriques2024}.}

\section{Conclusion}

In this work, ionization and thermalization by $\alpha$-particles and spontaneous-fission fragments in kilonova ejecta were studied self-consistently, including stopping powers, particle spectra, charge evolution, and interactions with bound and free electrons. Corresponding calculations for $\beta$-electrons were also performed to compare the ionization efficiencies of the different radioactive decay channels.

The main result is that the work per ion pair is remarkably robust across all considered decay products. Despite the large differences in particle mass, charge, and injection energy between $\beta$-electrons, $\alpha$-particles, and fission fragments, the resulting ionization efficiencies are found to be quantitatively very similar, particularly for heavy-element ejecta compositions relevant to kilonovae. This behaviour follows from the common collision physics governing both stopping and ionization, which causes the ratio between ionization and energy loss to remain approximately constant over the relevant energy range and ejecta compositions.

These results suggest that the late-time ionization state of kilonova ejecta depends only weakly on which radioactive decay channel dominates the heating, even in ejecta enriched by actinides where $\alpha$-decay and fission become important.

\section*{Acknowledgements}
We thank Dr Tanja Hinderer and the anonymous referee for useful discussion and comments.
This work  was supported by the JST FOREST Program (JPMJFR2136) and the JSPS Grant-in-Aid for Scientific Research (23H01169, 23H04900).
\section*{Data Availability}
The data underlying this article will be shared on reasonable request
to the corresponding authors.



\bibliographystyle{mnras}
\bibliography{example} 




\appendix

\section{Effects of excitation and Auger electrons on electron degradation spectrum}\label{app1}

In this work, we do not account for stopping due to excitation and for the contribution of Auger electrons in the electron degradation spectrum (equation \ref{eq:deg}). \textcolor{black}{To assess the impact of this simplification, we compute the degradation spectrum of $\beta$ electrons in Nd III medium with $\chi_e=0.1$, $1$, and $2$ including both effects.} The excitation cross sections and Auger rates are computed using \texttt{HULLAC}.

\begin{figure}
    \centering
    \includegraphics[width=\linewidth]{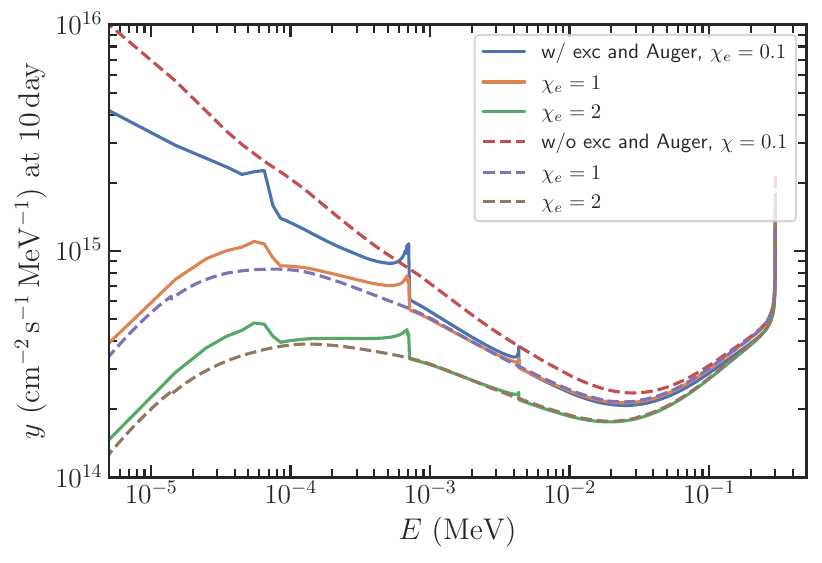}
    \caption{Electron degradation spectrum for Nd III medium including electron impact excitation and Auger electrons ({\it solid}) for electron fractions of $\chi_e=0.1$, 1, and 2. The spectra without excitation and Auger electrons are shown by {\it dashed} line. }
    \label{fig:app}
\end{figure}

\textcolor{black}{Figure \ref{fig:app} shows the degradation spectra computed with and without these effects.
For $\chi_e \geq 1$, we find that the spectra with and without excitation and Auger effects are identical down to $\sim 1$ keV. At lower energies, Auger electrons enhance the degradation spectrum by up to $\approx 10\%$.  Similarly, the work per ion pair is expected to increase by only $\mathcal{O}(10\%)$. Therefore, we conclude that neglecting excitation and Auger electrons does not significantly affect the ionization efficiencies in kilonova plasma. This can be understood from the contributions of different processes to the stopping power. For example, in the case of $\chi_e \gtrsim 1$,
ionization and Coulomb collision dominate the stopping power for $E\gtrsim 10\,{\rm keV}$ and $E\lesssim 10\,{\rm keV}$, respectively. As a result, the contribution of excitation to the stopping power is less than $10\%$ over the entire range of the degradation spectrum (see, e.g., figure 6 of \citealt{Andalman2026}).}

\textcolor{black}{On the contrary, for nearly neutral plasma $\chi_e\sim 0.1$, the excitation stopping power is important for $E\lesssim 10\,{\rm keV}$ because the stopping power due to Coulomb collision is rather weak. As a result, the electron fluxes without excitation are higher by a factor $\sim 1.2$ and $2$ at 1 keV and 10 eV, respectively. Therefore, the work per ion pair can be underestimated by up to a factor of $\sim 2$ when neglecting the excitation stopping. }

\section{Analytic Scaling for Work per Ion Pair}\label{app2}
We show a simple analytic estimate for the work per ion pair by considering how the ionization rate depends on the atomic structure of the stopping medium. For a given degradation spectrum, $y(E)$,
the production rate of electron-ion pairs is
\begin{align}
    \mathcal{N}_i \propto \sum_n \int_{I_n}^{E_0} dE \sigma_{{\rm ion},n}(E) {y}(E), 
\end{align}
where the summation is over atomic shells and $\sigma_{{\rm ion},n}(E)$ is the ionization cross section for a shell $n$ with $N_n$ electrons:
\begin{align}
    \sigma_{{\rm ion},n}(E) \propto \frac{N_n}{I_n} \frac{\ln(E/I_n)}{E}.
\end{align}
Therefore,
\begin{align}
    \mathcal{N}_i \propto \sum_n \frac{N_n}{I_n} \int_{I_n}^{E_0} dE \frac{{y}(E)\ln(E/I_n)}{E}.
\end{align}
The integral part is only weakly dependent on $I_n$ and the injection energy $E_0$, and thus, we obtain
\begin{align}
    \mathcal{N}_i \propto \sum_n \frac{N_n}{I_n}.
\end{align}
Using this, the work per ion pair is inversely proportional to the number of electrons weighted by the potential energies of shells 
\begin{align}
    {\rm WIP} \propto \left(\sum_n \frac{I_0}{I_n}N_n\right)^{-1}.\label{eq:scale}
\end{align}
This approximate scaling enables the computation of the work per ion pair for various ion species for a given $y(E)$.


\bsp	
\label{lastpage}
\end{document}